\documentclass[pra,twocolumn,groupedaddress,showpacs,amsmath,amssymb,floatfix]{revtex4}
\usepackage{graphicx}
\usepackage{dcolumn}
\usepackage{bm}
\def\btt#1{\texttt{\@backslashchar#1}}
\DeclareRobustCommand\bblash{\btt{\@backslashchar}}

\begin{document}

\title{Single-qubit rotations in 2D optical lattices with multi-qubit addressing}

\author{Jaewoo Joo$^1$} \email{jaewoo.joo@imperial.ac.uk}
\author{Yuan Liang Lim$^1$}
\author{Almut Beige$^2$}
\author{Peter L. Knight$^1$}

\affiliation{$^1$Blackett Laboratory, Imperial College London, Prince Consort Road, London, SW7 2BW, United Kingdom\\
$^2$The School of Physics and Astronomy, University of Leeds, Leeds, LS2 9JT, United Kingdom}

\date{\today}

\begin{abstract}
Optical lattices with one atom on each site and interacting via cold controlled collisions provide an efficient way to entangle a large number of qubits with high fidelity. It has already been demonstrated experimentally that this approach is especially suited for the generation of cluster states [O. Mandel {\em et al.}, Nature {\bf 425}, 937 (2003)] which reduce the resource requirement for quantum computing to the ability to perform single-qubit rotations and qubit read out. In this paper, we describe how to implement these rotations in 1D and 2D optical lattices without having to address the atoms individually with a laser field.
\end{abstract}

\maketitle

\section{Introduction} \label{Intro}

\begin{figure}[b]
\begin{minipage}{\columnwidth}
\begin{center}
\resizebox{\columnwidth}{!}{\rotatebox{0}{\includegraphics{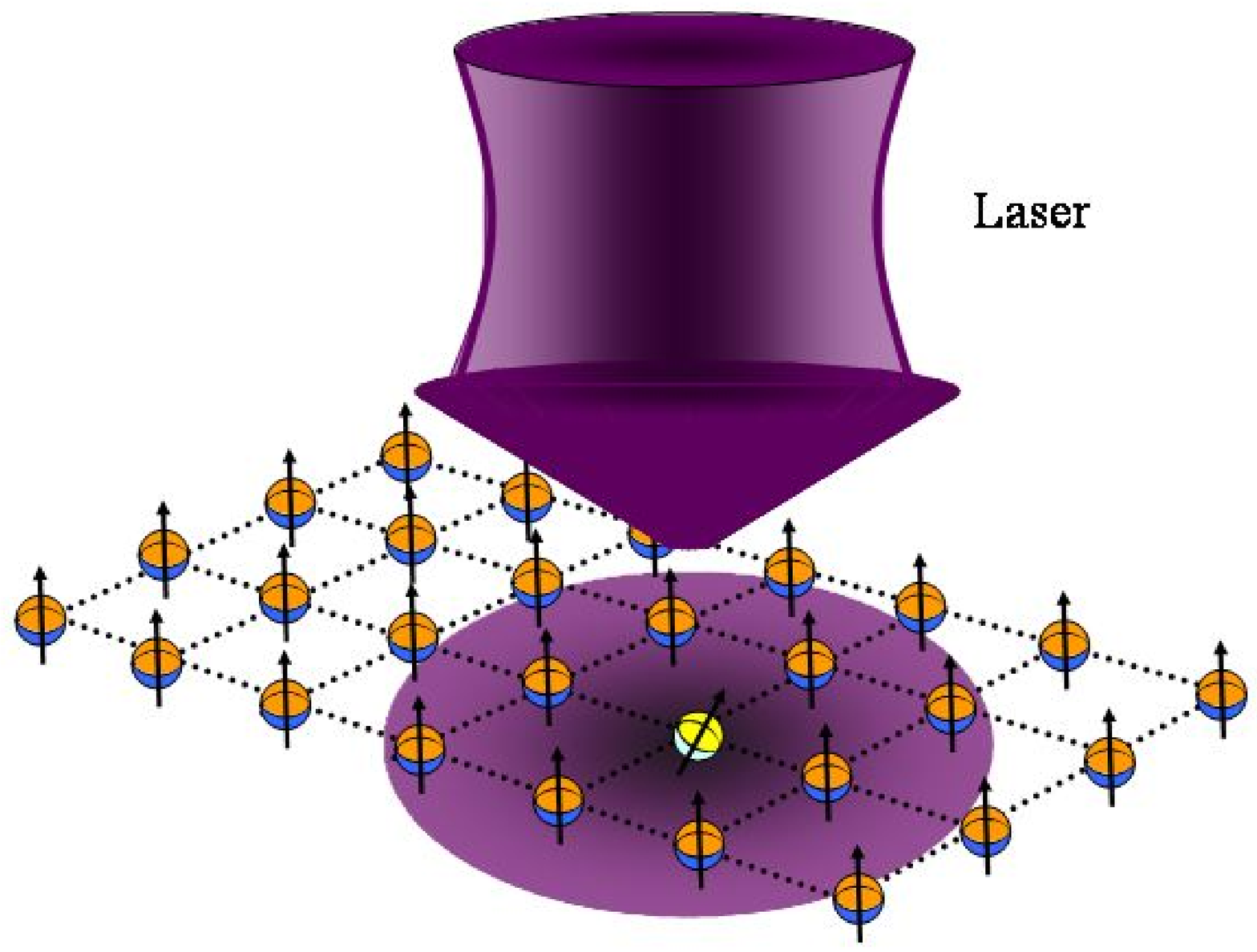}}}
\end{center}
\vspace{-1.5cm}
\caption{A 2D optical lattice structure with one atom per site in the presence of an optical laser field. Single-qubit addressing is limited by the fact that the size of the laser waist cannot be made small compared to the distance between neighbouring atoms.} \label{fig:2Dcluster}
\end{minipage}
\end{figure}

In recent years, much effort has been expended to find efficient ways to implement quantum computing in the laboratory \cite{rev}. Experiments demonstrating universal two-qubit gate operations using ion trap technology \cite{Blatt,Wineland} and photonic qubits \cite{Franson,Pan} have already been performed. Prototype algorithms operating on a few qubits, and basic quantum error correction have recently been demonstrated in the laboratory \cite{Chiaverini,Gulde,Steffen,Bhattacharya,Vandersypen}. However, the only physical setup which currently allows for the controlled generation and manipulation of entanglement of a {\em very large} number of qubits consists of atoms trapped inside an optical lattice, as shown in Figure \ref{fig:2Dcluster}. Such a lattice is formed by counter propagating laser fields producing an array of atomic traps \cite{Meacher}. Loading a lattice with a cloud of cold neutral atoms and performing a Mott transition results ideally in a situation with exactly one atom per site \cite{mott1,mott2}. 

The dependence of the trapping potential on the internal state of
the atoms (the qubits) allows to manipulate them efficiently. More concretely, switching the trapping
parameters \cite{Brennen,collision} or lowering the barrier
between sites \cite{Pachos03} results in the generation of a
controlled interaction between neighbouring qubits. In this way a
conditional phase shift is obtained which is especially suited to
generate {\em cluster states} \cite{cluster}. Cluster states are a
class of highly entangled many-qubit states. Their preparation
requires nothing else than the application of a controlled phase gate to all neighbouring sites in the optical
lattice, which can be realised within a few parallel steps. Mandel {\em et al.} \cite{Mandel} recently generated a cluster state and reported the observation of coherence between the wave packets of an atom delocalized over many sites. Theoretical proposals for the purification of cluster states can be found in the literature \cite{Duer,Kay2,rob}. 

In 2001, Briegel and Raussendorf showed that cluster states constitute a very useful resource for quantum computing \cite{oneway}. Once a cluster state has been built, a so-called one-way quantum computation can be performed to realise any possible quantum algorithm. It requires only single-qubit rotations,
\begin{eqnarray} \label{urot}
U_{\rm rot} (\xi, \phi) & \equiv& \cos \xi - {\rm i} \, \sin \xi \big( \, {\rm e}^{{\rm i} \phi} \, |0 \rangle \! \langle 1| + {\rm
H.c.} \, \big) \, , 
\end{eqnarray} 
and single-qubit read out, i.e.~measurements whether a qubit is in $|0 \rangle$ or $|1 \rangle$. Scalable fault-tolerant one-way computation with 2D and 3D cluster states is possible, provided the noise in the implementation is below a certain threshold \cite{nielsen,nielsen2}. For example, Raussendorf {\em et al.} \cite{rob2} recently introduced a fault-tolerant 3D cluster state quantum computer based on methods of topological quantum error correction and derived a threshold of $0.11 \% $ for each source in an error model with preparation, gate, storage and measurement errors. 

Moreover the actual implementation of one-way quantum computing in optical lattices remains experimentally challenging. The main problem arises from the fact that the typical distance between two lattice sites equals half
the wavelength of the trapping laser. The addressing and the trapping laser are in general of comparable optical wavelength. Consequently, single-qubit rotations cannot be realised as usual by applying
a single laser pulse. The minimum beam waist of the applied field is unavoidably
of similar size as the distance between neighbouring atoms. The
laser therefore not only affects the state of the target atom but
also the state of its neighbours (c.f.~Figure \ref{fig:2Dcluster}).
Several proposals addressing this problem and showing how to
realise single-qubit rotations without having to address the atoms
individually with a laser field have already been made
\cite{Friebel,Phillips,Wunderlich,Kay04,Calarco,Solano,You,Kay,benjamin}.

To improve the addressability of the qubits one can try to enlarge
the distance between neighbouring lattice sites by trapping the
atoms with a laser with a relatively long wavelength, such as a
$CO_2$ laser \cite{Friebel}, and by tilting the trapping laser
accordingly \cite{Phillips}. Mintert and Wunderlich
\cite{Wunderlich} proposed to overcome the qubit addressability
problem by using microwave transitions, whose degeneracy can be
lifted with the help of a strong inhomogenous magnetic field.
Furthermore, it has been proposed to implement single-qubit
rotations with the help of pointer atoms
\cite{Kay04,Calarco,Solano,You,Kay,benjamin}. To perform a single-qubit
rotation, a pointer atom in the form of a travelling address
``head" should be moved to the position of the respective atom
where it initiates the desired operation. Creating a pointer atom
either requires natural defects \cite{Solano} or the ability to
address a qubit individually at one point within the lattice
\cite{Kay04}. Schemes based on pointer atoms further require the
ability to transport atoms over many lattice sites without losing
coherence.

In this paper, we propose an alternative realisation of
single-qubit rotations without having to address atoms
individually with a laser field. We describe two different but
related approaches based on the experimentally more convenient
technique of {\em multi-qubit} addressing. The first approach is inspired by recent ion
trap experiments \cite{NIST96} and requires the simultaneous
excitation of the atoms by several laser fields. The directions of
the incoming laser fields are chosen such that the lasers
interfere classically at the different atomic positions. Within
the resulting fringe pattern, only the target atom sees a
non-vanishing Rabi frequency and undergoes a rotation. The second
approach is inspired by a recent paper by Pachos and Walther
\cite{Pachos02} and composite pulse techniques \cite{NMR}
typically used in NMR experiments. It requires the sequential
excitation of atoms with laser beams coming from different
directions. These directions are chosen such that any unwanted
rotation caused by one pulse is compensated by another
pulse. Implementing a single-qubit rotation is possible, since the dependence of the phase factor of the laser Rabi frequency on the position of the respective atom makes it possible to treat the target atom differently from the
surrounding qubits.

Suppose a laser field is applied with its centre focused on the target qubit in a 1D optical lattice, thereby unavoidably addressing $2N$ additional qubits. This means, the laser irradiates $N$ atoms on the left and $N$ atoms on the right
side of the target qubit. Performing a single-qubit rotation requires interfering $N+1$ laser fields. Alternatively, if
$N+1$ is a power of 2, i.e.
\begin{equation} \label{N}
N+1=2^L
\end{equation}
with $L$ being a positive integer, it can be realised using $N+1$ subsequent laser pulses. Both approaches are comparable in resources. Which one to favour depends on the available experimental tools.

This paper is organised as follows. In Section \ref{classic}, we discuss the possible implementation of single-qubit rotations using multi-qubit addressing and classical interference of laser beams in a 1D optical lattice. In Section \ref{Main1}, we describe the realisation of the same operation with sequential laser pulses. Section \ref{high-D} generalises our scheme to the more interesting case of 2D optical lattices, which are sufficient for the implementation of universal one-way quantum computation. Finally, we summarise our results in Section \ref{conc}.

\section{Single-qubit rotations in a 1D optical lattice via classical interference} \label{classic}

The realisation of a single-qubit
rotation on a single two-level atom requires the generation of the Hamiltonian
\begin{eqnarray} \label{H}
H  =  {\textstyle {1 \over 2}} \hbar \, \left( \, {\rm e}^{{\rm i} \varphi} \Omega \, |0 \rangle \langle 1| +   {\rm H.c.} \right) \, .
\end{eqnarray}
Here ${\rm e}^{{\rm i} \varphi} \Omega$ denotes an effective Rabi
frequency and we can assume that $\Omega$ is real without loss of generality.
Calculating the corresponding time evolution operator, we obtain
\begin{eqnarray} \label{U}
U (t_1,t_0) &=& \cos \Big( \int_0^{\Delta t} {\textstyle {{1 \over 2}}} \Omega(t) \,  {\rm d} t \Big) \big ( \, |0 \rangle \langle 0| + |1 \rangle \langle 1| \, \big) \nonumber \\
&& - {\rm i} \, \sin \Big( \int_0^{\Delta t} {\textstyle {{1 \over 2}}} \Omega(t) \,  {\rm d} t \Big) \big ( \, {\rm e}^{{\rm i} \varphi} \, |0 \rangle \langle 1| + {\rm H.c.} \, \big) \, , \nonumber \\
\end{eqnarray}
where $\Delta t = t_1 - t_0$. In case of a constant Rabi frequency ($\Omega(t) \equiv \Omega)$, the operator (\ref{U}) coincides with the single qubit rotation (\ref{urot}) if $\xi = {\textstyle {{1 \over 2}}} \Omega \Delta t$ and $\phi = \varphi$.
Choosing $\varphi$ and $\Delta t$ appropriately allows for the realisation of any single-qubit rotation of form (\ref{urot}).

\begin{figure}[b]
\centering
\begin{minipage}{\columnwidth}
\begin{center}
\resizebox{\columnwidth}{!}{\rotatebox{0}{\includegraphics{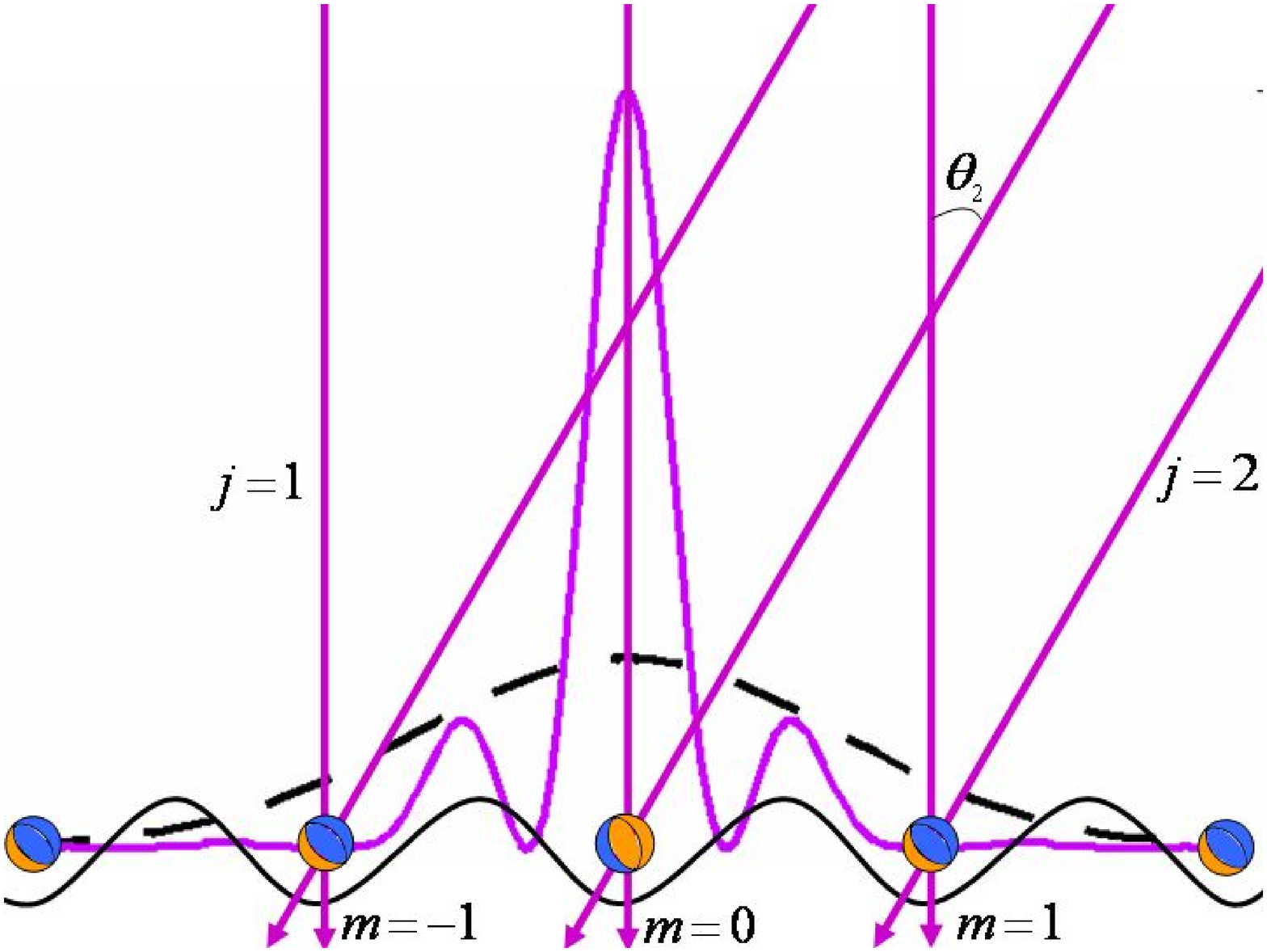}}}
\end{center}
\vspace{-0.1in} \caption{Classical interference of two identical
laser beams in the presence of an 1D optical lattice. If one laser field
approaches the setup from a direction perpendicular to the line
connecting the atoms while the second field is tilted by an angle
$\sin \theta_2 = \lambda_{\rm L}/\lambda_{\rm T}$, the setup can
be used to implement a single-qubit rotation with 3-qubit
addressing.} \label{fig:TiltLaser}
\end{minipage}
\end{figure}

When one tries to realise the Hamiltonian (\ref{H}) by focusing a
laser on an atom trapped inside a 1D optical lattice, it
unavoidably affects also $2N$ neighbouring qubits. In the
following we show how the above Hamiltonian (\ref{H}) can
nevertheless be realised with the help of classical
interference of several laser beams. Suppose diffractive optics
is used to divide the laser field into $N+1$ identical sub-beams
and the direction of the first beam $(j=1)$ is perpendicular to
the line connecting the atoms. The other beams $(j=2, \, ..., \,
N+1)$ are tilted with respect to this first one as shown in Figure
\ref{fig:TiltLaser} and focus also on the target atom. The
difference between the beams lies only in the direction in
which they approach the 1D lattice. Beam $j$
differs from the first beam by a rotation by an angle
$\theta_j$ around the target atom.

In the following, ${\rm e}^{{\rm i} \varphi_j (m)} \Omega_j (m)$
denotes the Rabi frequency of beam $j$ with respect to atom $m$.
We assume that $ \Omega_j (m)$ is real and choose the notation such that $m=\pm1, \,..., \, \pm N$ counts the distance of a qubit from the target qubit with $m=0$.
Geometrical considerations imply that the phase difference of the
same laser field between two neighbouring sites depends on $j$ but
not on $m$ (c.f.~Figure \ref{fig:TiltLaser}). We denote this phase
factor as $\vartheta_{j}$ and note that
\begin{eqnarray} \label{vartheta}
\vartheta_{j} = \varphi_{j} (m) - \varphi_{j}(m-1)\, .
\end{eqnarray}
Taking into account that $\varphi_j(0) = \varphi_1(0)$ for all
$j$, we find that
\begin{equation} \label{m}
\varphi_j(m) = m \, \vartheta_{j} + \varphi_1(0) \, .
\end{equation}
If the laser fields have a relatively broad Gaussian mode profile,
one can moreover assume that rotating a beam around the target
atom does not change $\Omega_j(m)$ and
\begin{eqnarray} \label{vartheta2}
\Omega_j(m) = \Omega_1(m)
\end{eqnarray}
for all $j$. Hence, the total Rabi frequency seen by qubit $m$ equals
\begin{eqnarray} \label{eq:classic01}
{\rm e}^{{\rm i} \varphi (m)} \Omega (m)
&\equiv & \sum_{j=1}^{N+1}  {\rm e}^{{\rm i} \varphi_j (m)} \Omega_j (m) \nonumber \\
&=&  {\rm e}^{{\rm i} \varphi_1 (0)} \Omega_1 (m) \sum_{j=1}^{N+1}  {\rm e}^{{\rm i} m \, \vartheta_j} \, . ~~~
\end{eqnarray}
We now need to find the angles $\vartheta_j$ such that only the target atom sees an effective Rabi frequency.

The right hand side of Eq.~(\ref{eq:classic01}) vanishes for all $m$ with $|m| \ge 1$ but not for $m=0$ if
\begin{eqnarray} \label{z}
\sum_{j=1}^{N+1} {\rm e}^{{\rm i} m \vartheta_j }  = 0 \, .
\end{eqnarray}
To fulfil this condition we assume that the $N+1$ laser fields are
tilted with respect to each other such that
\begin{eqnarray} \label{vari-theta}
\vartheta_j  = {2 \pi \, (j - 1) \over N+1} \, ,
\end{eqnarray}
since this implies
\begin{eqnarray} \label{z2}
\sum_{j=1}^{N+1} {\rm e}^{{\rm i} m \vartheta_j }
&=& \sum_{j=0}^N \left( {\rm e}^{ 2 \pi {\rm i} m / (N+1)} \right)^j \, .
\end{eqnarray}
This is indeed zero for all $m \neq 0$ since
\begin{eqnarray} \label{z3}
z^0 + z^1 + ... + z^N = 0
\end{eqnarray}
for all complex numbers $z$ with $z^{N+1} =1$ and $z \neq 1$. If Eq.~(\ref{vari-theta}) applies, all $N+1$ laser beams interfere destructively at the positions of the atoms with $m \neq 0$ but result in a single-qubit rotation on the target atom. Using again Eq.~(\ref{eq:classic01}) but with $m=0$ we find that
\begin{eqnarray} \label{eq:classic03}
{\rm e}^{{\rm i} \varphi (0)} \Omega (0) = (N+1) \,  {\rm e}^{{\rm i} \varphi_1 (0)} \Omega_1 (0) \, .
\end{eqnarray}
This is the effective Rabi frequency ${\rm e}^{{\rm i} \varphi} \Omega$ experienced by the qubit with $m=0$.

Using laser pulses of equal length and equal amplitude, as we propose here, $N+1$ is the minimum number of beams required to implement a single qubit rotation when addressing $2N+1$ atoms simultaneously. To implement the proposed scheme, one also needs to know how to
choose the tilting angles $\theta_j$ (c.f.~Figure \ref{fig:TiltLaser}). Suppose $\lambda_{\rm L}$ is the wavelength of the applied laser field and $\lambda_{\rm T}$ is the wavelength
of the trapping laser. The distance between two neighbouring atoms
is then given by ${1 \over 2} \lambda_{\rm T}$. Taking this into account, we
find
\begin{eqnarray} \label{anglecondition}
\sin \theta_j = {\vartheta_j \over \pi} \cdot {\lambda_{\rm L} \over \lambda_{\rm T}} \, .
\end{eqnarray}
The value of each $\vartheta_j$ is given in Eq.~(\ref{vari-theta}) for any given number $N$.

\subsection{3-qubit addressing $(N=1)$} \label{hallo}

For example, the implementation of a single-qubit rotation with 3-qubit addressing ($N=1$) requires the classical interference of two laser beams (c.f.~Figure \ref{fig:TiltLaser}). As suggested by Eq.~(\ref{vari-theta}), their wave vectors should be chosen such that $\vartheta_1 = 0$ and $\vartheta_{2} = \pi$. This can be achieved by splitting one laser beam into two equally strong beams and tilting one of them with respect to the other before shining them onto the atoms. As Eq.~(\ref{eq:classic03}) shows, the target atom experiences the effective Rabi frequency $2 \, {\rm e}^{{\rm i} \varphi_1 (0)} \Omega_1 (0)$ while the other two atoms remain in their initial state. 

\begin{figure}[t]
\centering
\begin{minipage}{\columnwidth}
\begin{center}
\resizebox{\columnwidth}{!}{\rotatebox{0}{\includegraphics{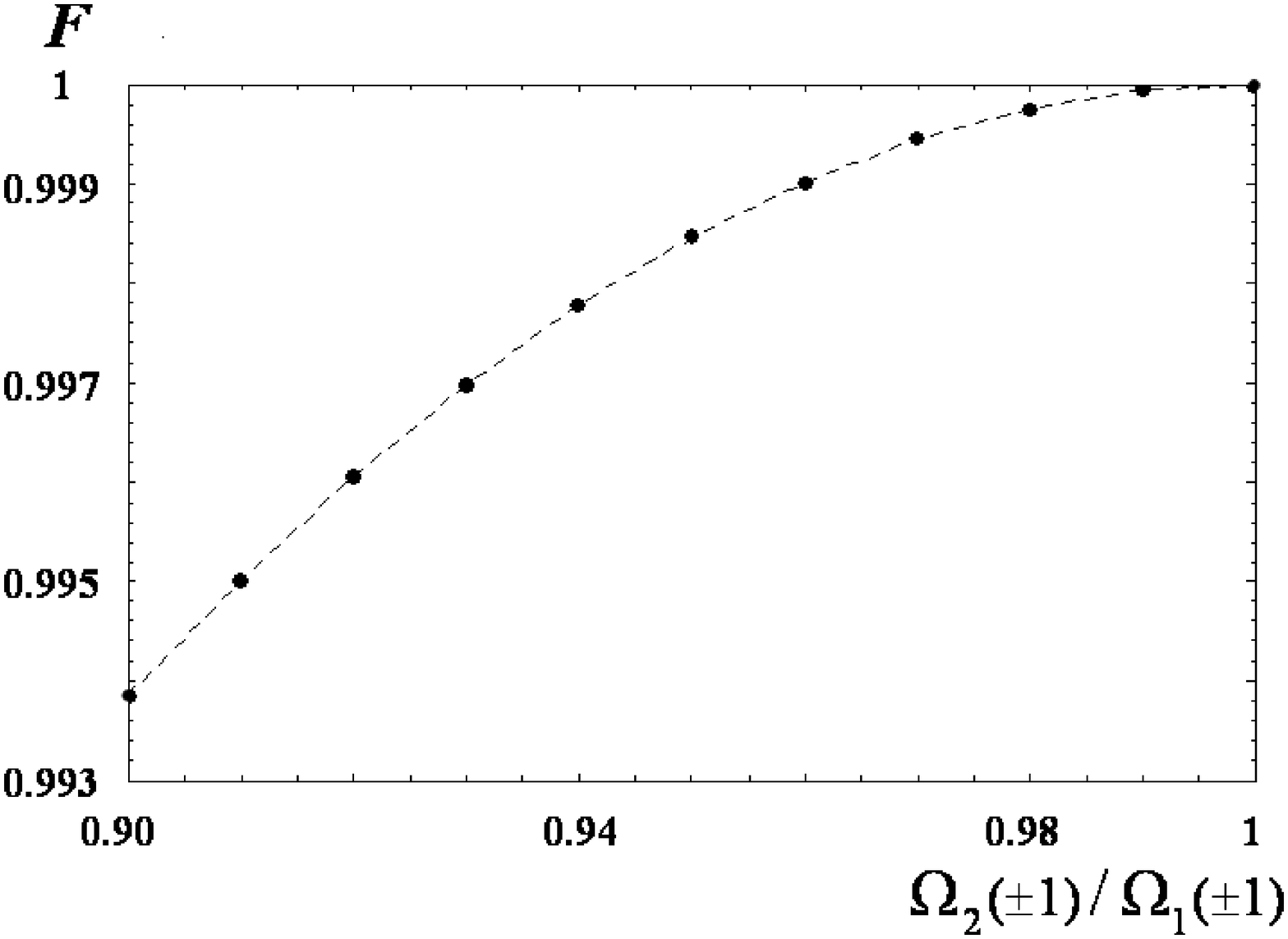}}}
\end{center} \caption{Fidelity $F$ of the final state of the atoms next to the target atom after the performance of a $\pi$-rotation on the target atom with 3-qubit addressing and two classically interfering laser fields. Different from Eq.~(\ref{vartheta2}), we assumed that $\Omega_2(\pm1) \neq \Omega_1(\pm1)$. Exactly the same decrease of the fidelity occurs, when the $\pi$ rotation on the target atom is performed using two successive laser pulses (c.f.~Section \ref{3qubit}). } \label{fig:error}
\end{minipage}
\end{figure}

In the derivation of this result we assumed that the same atom always sees the same Rabi frequency up to a phase factor, i.e.~$|\Omega_j(\pm1)|$ is for all $j$ the same (c.f.~Eq.~(\ref{vartheta2})). However, the tilting of a laser with a Gaussian mode profile around the target atom results unavoidably in a slight change of the laser amplitude and $\Omega_2 (\pm 1)$ is in general slightly smaller than $\Omega_1 (\pm 1)$. As a result, the atoms next to the target atom do not remain in their initial state. Figure \ref{fig:error} shows a lower bound for the overlap of the final state with the initial state for different values of $\Omega_2 (\pm 1)/\Omega_1 (\pm 1)$. In the derivation of this bound we considered the worst case scenario of the performance of a $\pi$-rotation on the target atom and assumed $\Omega_1(\pm1) = \Omega_1(0)$ although $|\Omega_1(\pm1)| < |\Omega_1(0)|$ in general. Even if the amplitude of the second beam differs by $10 \%$ from that of the first one, the fidelity with which the non-target atoms remain in their initial state is well above $0.99$. Errors below $0.1\%$ require that $\Omega_1 (\pm 1)$ and $\Omega_2 (\pm 1)$ differ by less than $4\%$. Errors arising from small phase or amplitude fluctuations of the laser Rabi frequency are expected to be even smaller than that. As one can see for example from Eq.~(\ref{U}), the time evolution of the atoms depends only on the integral over the Rabi frequency. Small fluctuations of the $\Omega$'s therefore average out.    

\subsection{Homogenous laser excitation $(N=\infty)$}

Implementing a single-qubit rotation on a
single atom inside a 1D optical lattice is even possible with a
laser beam with an infinite waist $(N \to \infty)$. To do so, the
beam should ideally be split in infinitely many sub-beams of equal intensity whose
wave vectors equally cover all possible incoming directions. In
position space the beams interfere such that they create an
electric field with a $\delta$-function like intensity peak at the position
of the target atom and zero amplitude for all other qubits within
the lattice.

\section{Single-qubit rotations in a 1D optical lattice via sequential laser pulses} \label{Main1}

Instead of applying $N+1$ laser fields simultaneously, a
single-qubit rotation can also be realised via the subsequent
application of $N+1$ laser pulses of length $\Delta t$. In the
following, we describe such a sequence which, even in the case of
$2N+1$ qubit addressing, results in the rotation of the center
atom but has no effect on the state of all other atoms. The
solution we present here is inspired by the mechanism used in the
previous section and works if $N$ can be written as in Eq.~(\ref{N})
with $L$ being an integer. The use of $N+1 = 2^L$ steps will allow
us to find a scheme in which the effective operation of a group of
laser pulses can always be undone by a subsequent group of the
same number of pulses for all qubits with $m \neq 0$.

In the previous section, the state of any atom with $m \neq 0$ does not change effectively, as long as the sum over all its Rabi frequencies (\ref{eq:classic01}) equals zero.
 If the same atom experiences instead a sequence of laser fields, then each field rotates the qubit state independently on the Bloch sphere.
The final operation on the
state of atom $m$ is given by
\begin{eqnarray} \label{eq:TotalLaser}
U_{m}(t_{N+1},t_{0}) &=& \prod_{k=1}^{N+1} {U}_{m} (t_{k},t_{k-1}) \, ,
\end{eqnarray}
where $t_{k-1}$ and $t_k$ are the starting and finishing time of
pulse $k$. Note that the time evolution operator
(\ref{eq:TotalLaser}) depends on the {\em exact order} in which
the fields are applied and the Rabi frequency used in each step.
The reason is that the commutator of two subsequent time evolution
operators
\begin{eqnarray} \label{commutation}
~~ && \hspace*{-1.2cm} [{U}_{m} (t_{k+1},t_{k})\, , {U}_{m} (t_{k},t_{k-1})]  \nonumber \\
&= & \big( \, {\rm e}^{{\rm i} ( \varphi_{k+1} (m) - \varphi_{k} (m))} - {\rm e}^{- {\rm i} ( \varphi_{k+1} (m) - \varphi_{k} (m))} \, \big) \nonumber \\
&& \times \sin^2 \big( {\textstyle {1 \over 2}} \Omega_k (m)  \Delta t \big) \, \big( \, |0 \rangle_{mm} \langle 0|
- |1 \rangle_{mm} \langle 1| \, \big) \nonumber \\
\end{eqnarray}
is in general not zero.

We consider the same Rabi frequencies and laser beams as in Section \ref{classic} but instead of applying all of them simultaneously, they should now be applied in a certain order. In the following we denote the Rabi frequency of the pulse applied in step $k$ and with respect to atom $m$ as ${\rm e}^{{\rm i} \tilde \varphi_k (m) } \tilde \Omega_k (m)$ and assume
\begin{eqnarray} \label{pi}
{\rm e}^{{\rm i} \tilde \varphi_k (m) } \tilde \Omega_k (m)
= {\rm e}^{{\rm i} \varphi_j (m) } \Omega_j (m)
\end{eqnarray}
with
\begin{eqnarray} \label{pi2}
j = \sigma (k) \, .
\end{eqnarray}
Here $\sigma$ is the permutation of the numbers $k= 1, \, 2, \, ...,\, N+1$. It tells us which beam $j$ (as described in the previous section) should be used in the $k$-th pulse of the sequence.

In this section we extensively employ the fact that two
subsequent laser fields, whose phases $\tilde \varphi_k(m)$ differ
by $\pm \pi$, have no effect on the state of qubit $m$. Let us
first consider, for example, the case of a pulse $k$ followed by a
pulse $k+1$ with
\begin{eqnarray} \label{cancel}
\tilde \varphi_{k+1}(m) = \tilde \varphi_{k}(m) \pm \pi \, .
\end{eqnarray}
Using Eq.~(\ref{U}) and taking Eq.~(\ref{vartheta2}) into account we find indeed that the corresponding
time evolution for the time interval $(t_{k+1},t_{k-1})$ equals
\begin{eqnarray} \label{cancel2}
U_m (t_{k+1},t_{k}) \, U_m (t_k,t_{k-1}) = {\bf 1} \, .
\end{eqnarray}
The effect of the first beam is {\em cancelled} by the effect of the second beam. For completeness we remark that for
\begin{eqnarray} \label{add}
\tilde \varphi_{k+1}(m) = \tilde \varphi_{k}(m)
\end{eqnarray}
one has
\begin{eqnarray} \label{add2}
U_m (t_{k+1},t_{k}) \, U_m (t_k,t_{k-1}) = \big[ \, U_m (t_k,t_{k-1}) \, \big]^2 \, .
\end{eqnarray}
If the Rabi frequencies of two subsequent pulses are the same, the respective time evolutions add up.

\subsection{3-qubit addressing $(N=1)$} \label{3qubit}

The simplest case is the realisation of a
single-qubit rotation using 3-qubit addressing. This requires only
two laser fields. Suppose the wave vector of the first laser is
perpendicular to the line connecting the atoms and $\tilde
\varphi_1 (m)$ is for all $m$ the same. To cancel the effect of
the first step on the states of the two outer qubits, the phase of the
second pulse should be $\tilde \varphi_2 (m) =  \tilde \varphi_1
(m) + \pi$ for $m=\pm 1$. This can be realised by simply tilting
the laser field used in the first step around the target atom by
the angle $\theta_2$ given in Eq.~(\ref{anglecondition}). From
Eqs.~(\ref{cancel2}) and (\ref{add2}) we see that the time
evolution of the atoms is given by
\begin{eqnarray} \label{eq:3qubit001}
{U}_{0}(t_2,t_0) &=& \big[ \, {U}_{0}(t_1,t_0) \, \big]^2 \, , \nonumber \\
{U}_{\pm1}(t_2,t_0) &=& {\bf 1} \, ,
\end{eqnarray}
which describes a single-qubit rotation on the target qubit while the states of the other two qubits do not change.

Small deviations from this prediction occur, if the the atoms next to the target atom see a different Rabi frequency 
in both laser pulses and $\Omega_2(\pm 1)$ differs from $\Omega_1(\pm 1)$ by a few percent. Different from Eq.~(\ref{vartheta2}), this is the case if the same laser is used in both steps up to a rotation by an angle $\theta_2$ (c.f.~Eq.~(\ref{anglecondition})). Figure \ref{fig:error} gives a lower bound for the fidelity and shows how well the state of the non-target atoms $(m=\pm 1)$ coincides with their initial state. As in Section \ref{hallo}, we consider the worst case scenario of the performance of a $\pi$ rotation on the target qubit and assume $\Omega_1(\pm 1) = \Omega_1(0)$ although $|\Omega_1(\pm 1)| < |\Omega_1(0)|$ in general. The errors arising in the case of sequential laser addressing of the atoms do not differ from the ones observed in Section \ref{hallo} and shown in Figure \ref{fig:error}.

\subsection{7-qubit addressing $(N=3)$} \label{7qubit_0}

This subsection illustrates our main ideas
for cancelling the effect of laser pulses for several
atoms inside a 1D optical lattice. We discuss
the realisation of a single-qubit rotation with 7-qubit addressing
using four subsequently applied laser pulses. Each step $k$ requires
a laser beam rotated by an angle $\theta_{\sigma(k)}$ (c.f.~Eq.~(\ref{anglecondition})) around the target atom. Since the target atom experiences the same Rabi frequency
\begin{eqnarray} \label{xxxx}
{\rm e}^{{\rm i} \tilde \varphi_k (0) } \tilde \Omega_k (0) = {\rm e}^{{\rm i} \varphi_1 (0)} \Omega_1 (0)
\end{eqnarray}
in each step $k$, the operation  performed on qubit $m=0$ equals
\begin{eqnarray}
U_0 (t_4,t_0) = \big[ \, U_0 (t_1,t_0) \, \big]^4  \, .
\end{eqnarray}
This time evolution allows the realisation of any single-qubit
rotation (\ref{urot}) if the Rabi frequency ${\rm e}^{{\rm i} \varphi_1 (0) }
\Omega_1 (0)$ is chosen accordingly.

\begin{figure}[b]
\centering
\begin{minipage}{\columnwidth}
\begin{center}
\resizebox{\columnwidth}{!}{\rotatebox{0}{\includegraphics{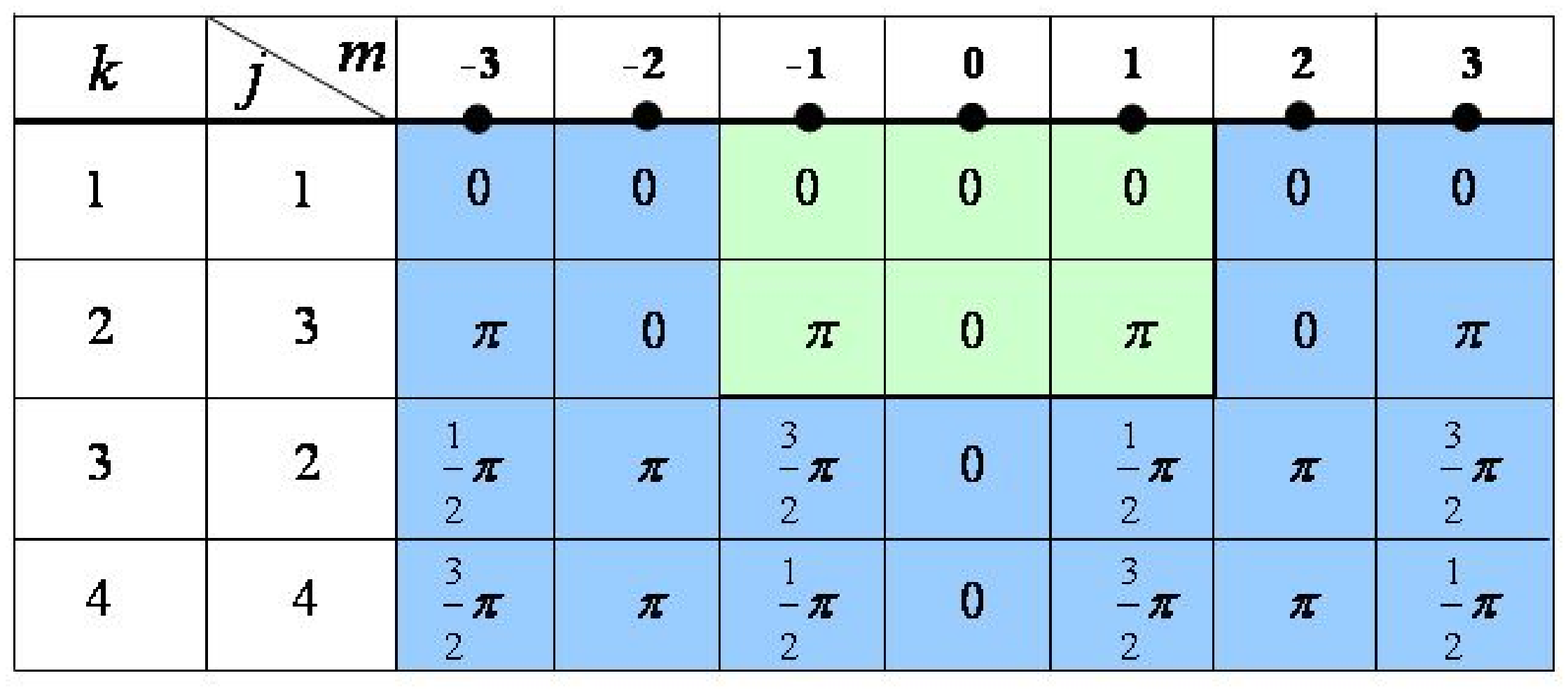}}}
\end{center}
\vspace*{-0.5cm} \caption{A possible sequence of laser pulses and the respective phases $\tilde \varphi_k(m) - \varphi_1(0)$ of the corresponding Rabi frequencies as a function of
$k$ and $m$ for the realisation of a single qubit rotation with 7-qubit addressing using four successive laser pulses.} \label{7qubit}
\end{minipage}
\end{figure}

Let us now consider the outer qubits. The wave vectors of the
laser pulses in steps 2, 3 and 4 could, for example, be chosen
such that
\begin{eqnarray} \label{seq}
\sigma(2) = 3 \, , ~~ \sigma(3) = 2 ~~ {\rm and} ~~ \sigma(4) = 4 \, .
\end{eqnarray}
This implies
\begin{eqnarray} \label{xxx}
&& \tilde \varphi_2 (m) - \tilde \varphi_1(0) = m \, \vartheta_3 = m \, \pi \, \nonumber \\
&& \tilde \varphi_3 (m) - \tilde \varphi_1(0) = m \, \vartheta_2 = {\textstyle {1 \over 2}} m \, \pi \, \nonumber \\
&& \tilde \varphi_4 (m) - \tilde \varphi_1(0) = m \, \vartheta_4 =
{\textstyle {3 \over 2}} m \, \pi
\end{eqnarray}
with the $\vartheta_{j}$'s as in Eq.~(\ref{vartheta}). The
right hand sides of these equations can be
found in the table shown in Figure \ref{7qubit}. There is no need
to know the concrete size of $\tilde \Omega_k(m)$ since we assume
that the amplitude of the applied fields is in each step the same (c.f.~Eq.~(\ref{vari-theta})).

Let us first have a closer look at the atoms $|m|=1$. We note that
the sequence (\ref{seq}) has been chosen such that the phase
factor $\tilde \varphi_k (\pm 1)$ of a step with an odd $k$ is
always followed by a step with a phase factor $\tilde \varphi_{k+1} (\pm 1)$ with
\begin{eqnarray} \label{rel}
\tilde \varphi_{k+1}(\pm 1) - \tilde \varphi_k (\pm 1) &=& \pi
\end{eqnarray}
up to multiples of $2 \pi$ (c.f.~Eq.~(\ref{cancel})). The rotation performed on the qubits
during the first laser pulse is therefore undone in the second step and the
rotation performed in step 3 is undone in step 4. Consequently,
the time evolution operator $U_{\pm 1}(t_4,t_0)$ is the identity operator
and the qubits $|m|=1$ do not change (c.f.~Eq.~(\ref{cancel2})).

The relation (\ref{rel}) also determines the relative phase
difference of the Rabi frequencies seen by the qubits with $|m| >
1$ in step $k$ and $k+1$. Using geometrical considerations (or
Eq.~(\ref{xxx})) one can show that
\begin{eqnarray} \label{rel2}
\tilde \varphi_{k+1}(m) - \tilde \varphi_k (m) &=& m \, \pi \, ,
\end{eqnarray}
for any odd number $k$. For the qubits with $|m| = 3$, the result on
the right hand side differs from $\pi$ only by a multiple of $2
\pi$ (c.f.~also Figure \ref{7qubit}). Consequently, $U_{\pm 3}(t_4,t_0)$ is the
identity operator and the states of the
two outer qubits with $|m| = 3$ at the end of the laser pulse sequence
is the same as in the beginning.

For the case $|m|=2$, Eq.~(\ref{rel2}) shows that the atoms see
the same Rabi frequencies between step $k$ and $k+1$ $(k=1, 3)$. The cancellation of the rotations performed in the
first two steps therefore requires that the rotation in the last
two steps are opposite to the rotation in the first two. This is the reason that we need at least four pulses and fixes the choice of the phase factor $\tilde \varphi_3 (1)$. It
should differ from $\tilde \varphi_1 (1)$ by ${1
\over 2} \pi$ as shown in Figure \ref{7qubit}. The four laser pulses then have no effect on the
atoms with $|m|=2$.

\subsection{$(2N+1)$-qubit addressing} \label{general}

In this subsection we describe the realisation of a single qubit rotation on the target qubit for the general case of $(2N+1)$-qubit addressing via subsequent laser pulses. This is possible if $N+1 = 2^L$ (c.f.~Eq.~(\ref{N})) and requires the application of the $N+1$ laser introduced in Section \ref{classic} in a certain order. According to Eq.~(\ref{m}), the target atom with $m=0$ experiences the same Rabi frequency ${\rm e}^{{\rm i} \varphi_1 (0) } \Omega_1 (0)$ in each step and the
total unitary operation is given by
\begin{eqnarray} \label{sumup}
U_0 (t_{N+1},t_0) = \big[ \, U_0 (t_1,t_0) \, \big]^{N+1} \, .
\end{eqnarray}
Here we present an algorithm that can be used to find a possible order in which to apply the beams such that the states of the outer qubits with $m \neq 0$ do not change. We use the same notation as in the previous subsection and the atoms see the laser beam $j = \sigma(k)$ with the Rabi frequencies (\ref{pi}) in step $k$.

The basic idea is to group the beams with the same phase factor $\varphi_j(m)$ together and arrange these groups such that they are always followed or succeeded by another group of laser pulses with a $\pi$-phase difference in their Rabi frequency. To do so, we should start by examining the middle outer qubits for which the Rabi frequencies have only two different phase factors. In the previous subsection, this was the case for $|m|=2$. In general, this applies for the qubits with
\begin{equation} \label{m0}
|m| = {\textstyle {1 \over 2}} (N+1) = 2^{L-1} \, .
\end{equation}
If the difference between phase factors seen by qubit 1 in step $k$ and step $l$ equals $x$,
\begin{equation} \label{a}
\tilde \varphi_k(1) - \tilde \varphi_l (1) = x  \, ,
\end{equation}
then the geometrical considerations that lead to Eq.~(\ref{m}) imply
\begin{equation} \label{b}
\tilde \varphi_k(m) - \tilde \varphi_l (m) = m \, x  \, .
\end{equation}
The smallest possible value of $x$ for $k \neq l$ is $\vartheta_2 = 2\pi/(N+1)$, in which case the right hand side of this equation becomes $\pi$. All other phase differences are multiples of this since $\vartheta_j$ is a multiple of $\vartheta_2$ (c.f.~Eq.~(\ref{vari-theta})). The angles $\varphi_j(m)$ equal therefore either $0$ or $\pi$. The first step in finding a possible permutation (\ref{pi2}) is to arrange the beams in an order such that $\tilde \varphi_k (m) = 0$ for $k = 1, \, ..., \, {\textstyle {1 \over 2}}(N+1)$
and $\tilde \varphi_k (m) = \pi$ for all other $k$.

\begin{figure}[b]
\centering
\begin{minipage}{\columnwidth}
\begin{center}
\resizebox{\columnwidth}{!}{\rotatebox{0}{\includegraphics{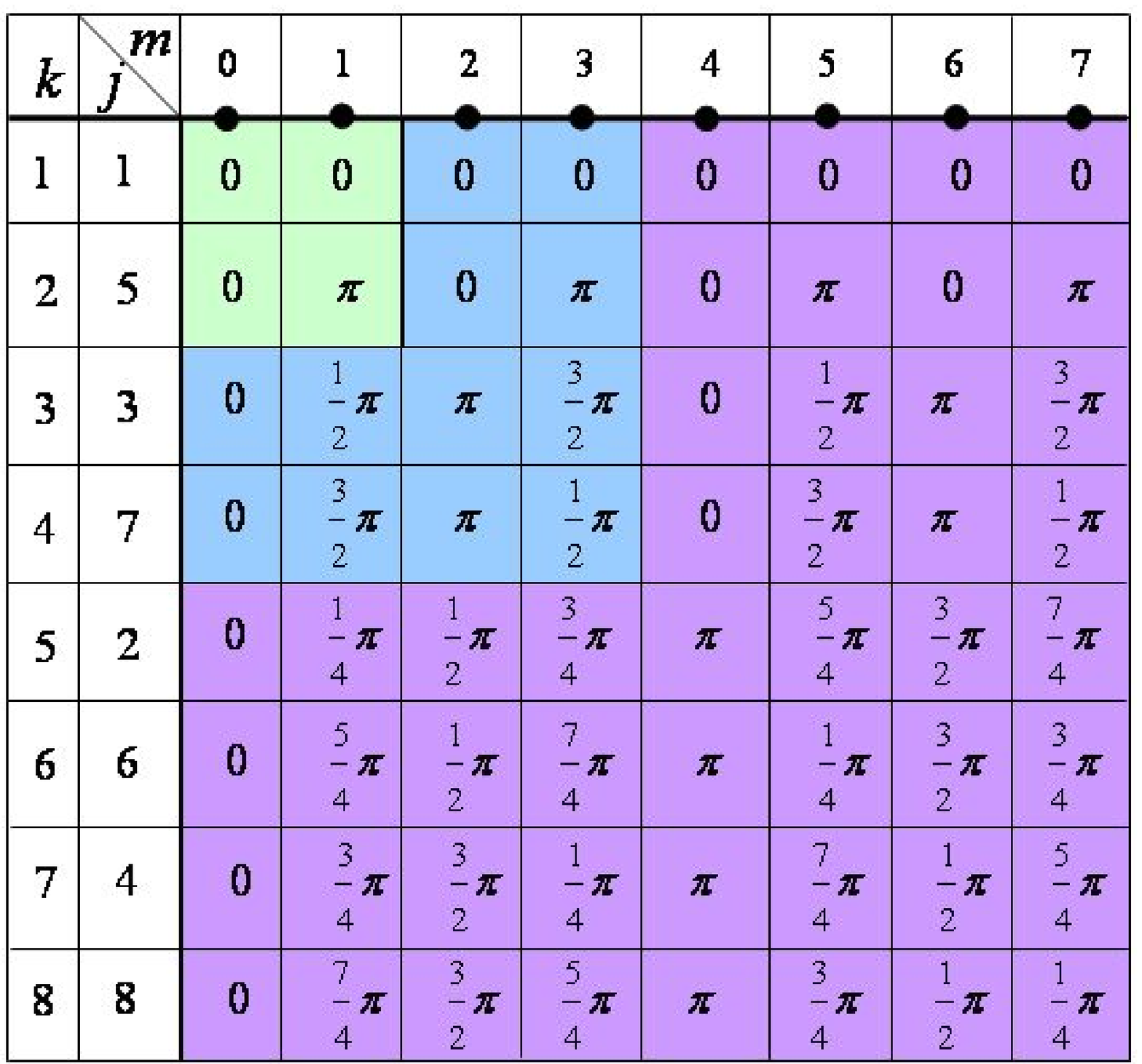}}}
\end{center}
\vspace{-0.1in}  \caption{A possible sequence of laser pulses and the respective phases $\tilde \varphi_k(m) - \varphi_1(0)$ of the corresponding Rabi frequencies as a function of
$k$ and $m$ for the realisation of a single qubit rotation with 15-qubit addressing using eight successive laser pulses.} \label{15qubitC}
\end{minipage}
\end{figure}

At this stage, we can arbitrarily rearrange the order of the first ${1 \over 2}(N+1)$ beams and the order of the following beams, respectively, without changing the fact that the states of the middle outer atoms do not change. In the second step, we use this to achieve a cancellation of rotations at the
qubits with
\begin{equation}
|m| = {\textstyle {1 \over 4}} (N+1) = 2^{L-2} \, .
\end{equation}
Since $m$ is now half the size of $m$ in Eq.~(\ref{m0}), also the phase
factors $\tilde \varphi_k(m)$ and $\tilde \varphi_l(m)$ in Eq.~(\ref{b})
might now differ by an angle half the size of what it was before. If the
phase difference was 0 (or $\pi$, respectively) for the qubits (\ref{m0}),
as it applies for the first group of beams, it now equals $0$ or $\pi$. We
might therefore have to change the order of the beams. As before, we group
the beams with the same phase factor together and follow them by a sequence
of laser pulses whose Rabi frequency has a $\pi$-phase difference. For
example, the first ${1
\over 4}(N+1)$ beams could all have the phase factor 0, the next set could
have $\pi$, the third set could have ${3 \over 2} \pi$ followed by beams
with $\tilde \varphi_k(m)= {1 \over 2} \pi$ (c.f.~Figure \ref{7qubit}, case
$|m|=1$).

In step $s$ of the algorithm to find a suitable permutation $\sigma(k)$, we consider the two qubits with
\begin{equation} \label{s}
|m| = 2^{L-s}
\end{equation}
and calculate the phase factors $\tilde \varphi_k(m)$ seen in each step $k$. If necessary, we change the order of some of the laser beams. Finally, there should be $2^s$ subgroups of subsequent laser pulses. Each group contains $2^{L-s}$ beams with the same phase factor $\tilde \varphi_k(m)$. It is always possible to arrange the beams in such an order that each group is followed or succeeded by a group of laser beams with a $\pi$ phase difference.

For example, in the final step ($s=L$) we ensure that the qubits with $|m|=1$ do not change in time. From Eq.~(\ref{m}) we see that
\begin{eqnarray} \label{middle04}
\varphi_{j} (\pm 1) - \varphi_1 (\pm 1) &=& \pm \, { 2 \pi \, (j - 1) \over N+1}.
\end{eqnarray}
The Rabi frequencies of all $N+1$ subsequent laser beams have different phase factors. However, Eq.~(\ref{m}) assures that they can be arranged in pairs such that each beam used in a step with an odd number $k$ is followed by beam with a $\pi$-phase difference in its Rabi frequency. Therefore, the qubits next to the target qubit return to their initial state every two steps.

The only thing left to show is that a cancellation of the laser rotations at qubits with $m$ as in Eq.~(\ref{s}) also automatically assures no effect of the applied laser pulses on all the remaining qubits.
These remaining qubits are all characterised by an $m$ of the form
\begin{equation} \label{ms}
|m| = (2n + 1) \cdot 2^{L-s}
\end{equation}
with $n$ being a positive integer, which is a multiple of the $m$ of a qubit for which a cancellation occurs. Using Eq.~(\ref{m}) one can show that
\begin{equation}
\tilde \varphi_k(m) - \tilde \varphi_l (m) = (2n + 1) \big[ \, \tilde \varphi_k \big(2^{L-s} \big) - \tilde \varphi_l \big( 2^{L-s} \big) \, \big] \, .
\end{equation}
If the phase difference in the squared brackets equals $\pi$, then
automatically also the effective phase difference on the left side becomes
$\pi$. If the phase difference on the right hand side vanishes, so does the
phase difference on the left. Therefore, effectively no time evolution takes
place for all qubits with an $m$  as in Eq.~(\ref{ms}) for exactly the same
reason as the one described in Section \ref{7qubit_0}. To illustrate this,
Figure \ref{15qubitC} shows a possible sequence of laser pulses and the
respective phases $\tilde \varphi_k(m) - \tilde \varphi_1(0)$ of the
corresponding Rabi frequencies for the realisation of a single qubit
rotation with 15-qubit addressing using eight successive laser pulses.

\section{Single-qubit rotations in 2D optical lattices} \label{high-D}

\begin{figure}[b]
\centering
\begin{minipage}{\columnwidth}
\begin{center}
\resizebox{\columnwidth}{!}{\rotatebox{0}{\includegraphics{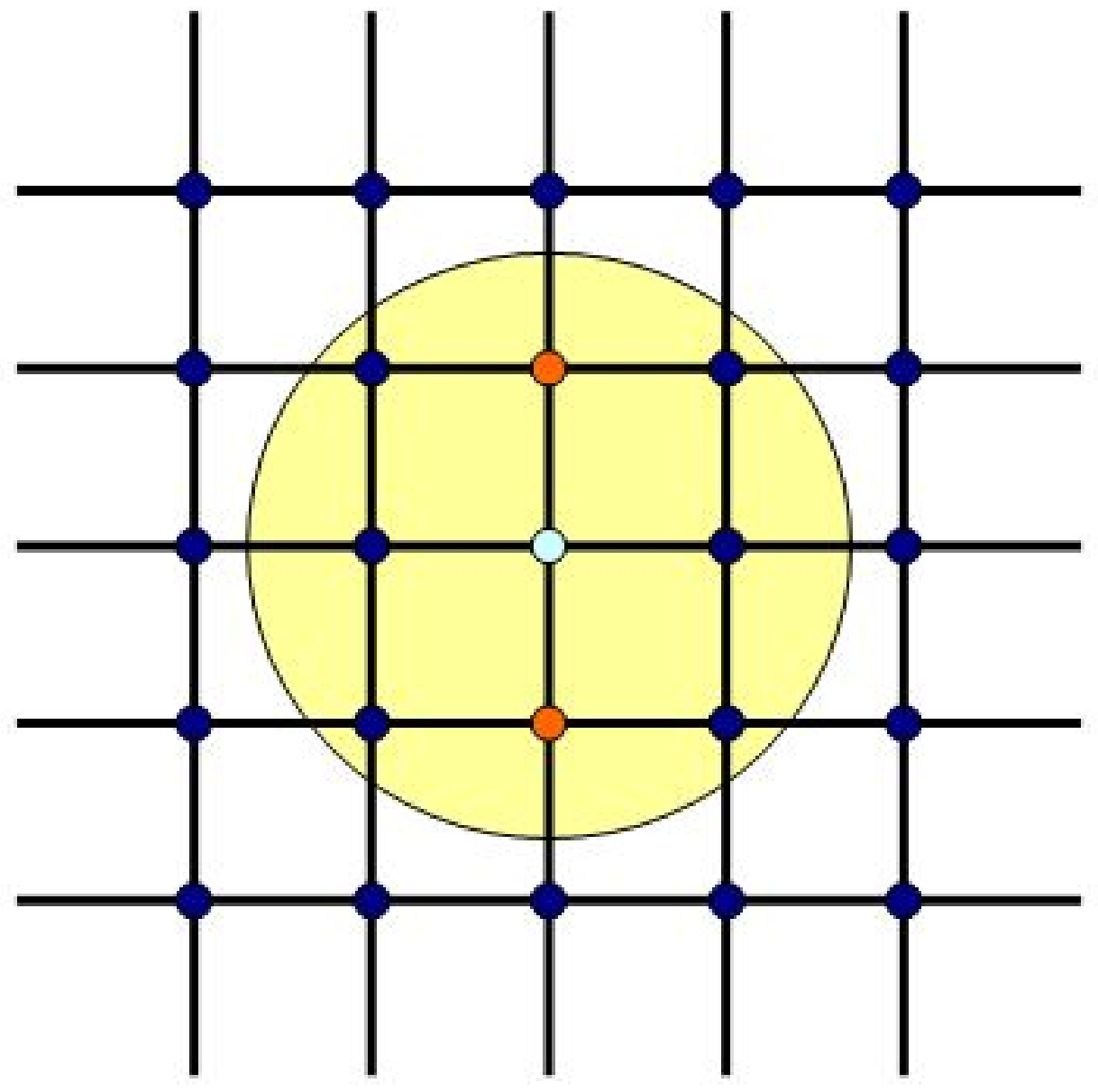}}}
\end{center}
\vspace{-0.1in} \caption{Atoms inside a 2D optical lattice structure addressed by a laser beam which simultaneously addresses three atoms in the same row. All together, the laser affects nine qubits.}
\label{fig:2D}
\end{minipage}
\end{figure}

In the following, we discuss ways to perform single-qubit rotations in 2D optical lattices with multi-qubit addressing. Suppose a laser beam of a certain size would excite $2N+1$, if the target atom is trapped inside a 1D lattice. The same laser then excites $(2N+1)^2$ atoms, when addressing a target atom in a 2D lattice. To change only the state of one atom, we make use of a technique that has been called the ``hiding of qubits'' in recent ion-trap experiments \cite{Roos}. It requires auxiliary levels and aims at transferring a subset of qubits, including the target qubit, into another Hilbert space, where the target atom occurs trapped in a 1D lattice. Single qubit rotations within this 1D lattice can be performed as described in Sections \ref{classic} and \ref{Main1}.

As an example, let us first consider the case in which a laser beam would address $N=3$ qubits in a 1D lattice. In a 2D lattice, such a laser beam excites nine atoms, as shown in Figure \ref{fig:2D}. The realisation of a single-qubit rotation now requires three steps. $(i)$ First a set of interfering laser fields or subsequently applied laser pulses should select a 1D chain of three atoms using the same techniques as described in Section \ref{hallo} or \ref{3qubit}. They should be applied such that the operation  
\begin{equation} \label{op}
|0 \rangle \leftrightarrow |0' \rangle ~~ {\rm and} ~~ |1 \rangle \leftrightarrow |1' \rangle 
\end{equation}
 is performed on a line of three atoms. $(ii)$ Afterwards, the central qubit among the three qubits (i.e.~the target qubit) can be rotated within two steps without affecting the state of the others by performing the desired rotation (\ref{urot}) between the states $|0' \rangle$ and $|1'\rangle$. $(iii)$ Finally, the operation (\ref{op}) needs to be undone by repeating step $(i)$. The generalisation of this idea to the case $N>1$ is straightforward.

\section{Conclusions} \label{conc}

We described two schemes for the realisation of single-qubit rotations in a 1D optical lattice with multi-qubit addressing. The first method requires the classical interference of $N+1$ laser beams, when exciting one central qubit and $2N$ additional ones. The angles between the beams should be chosen such that only the target atom sees a non-vanishing Rabi frequency. The second scheme utilises the same $N+1$ laser beams but instead of applying them all simultaneously the beams excite the atoms successively. This method can be used if $N+1 = 2^L$ with $L$ being an integer. The order in which the beams are applied is crucial for the scheme to work because the time evolution operators corresponding to different steps do not always commute with each other. Since both methods are similarly demanding in resources, which method to choose should depend on the available experimental tools. Realising single-qubit rotations in 2D optical lattice structures is possible after reducing the problem to the 1D case. This can be done by transferring a row of qubits, including the target qubit, into an auxiliary Hilbert space.  

In Sections \ref{classic} and \ref{Main1} we assumed that the same atom always sees the same Rabi frequency up to a phase factor. However, tilting a laser with a Gaussian mode profile around the target atom results in general in a slight change of this Rabi frequency. The result is that atoms other than the target qubit do not exactly remain in or return into their initial state. For $N=1$ (3-qubit addressing), we showed that the non-target atoms keep their initial state with a fidelity above $0.999$ as long as their Rabi frequency changes less than $4 \%$. For $N>1$ we expect smaller errors, since the tilting angles of the beams are in general smaller than in the $N=1$ case (c.f. Eqs.~(\ref{vari-theta})~and~(\ref{anglecondition})) while the duration of the gate operation is the same. Moreover, most atoms experience Rabi frequencies much smaller than the one seen by the target atom (c.f.~Figure \ref{fig:TiltLaser}). Figure \ref{fig:error} therefore indicates a lower bound for the fidelity with which the non-target atoms keep their initial state for all $N \ge 1$. Errors arising from small phase or amplitude fluctuations of the laser Rabi frequency are expected to be negligible. We are therefore optimistic that the proposed scheme has applications in quantum computing, although this might require a very precise control over the laser parameters. The exact threshold for scalable fault-tolerant one-way computation with optical lattice 2D cluster states is not yet known. Estimates of similar thresholds can be found in Refs.~\cite{nielsen,nielsen2,rob2}. \\

\noindent {\em Acknowledgment.} We thank Michael Trupke, Jeremy Metz, and Mark Tame for helpful discussions. J. J. is supported by an IT Scholarship from the Ministry of Information and Communication, Republic of Korea and the Overseas Research Student Award Program. Y. L. L. acknowledges the DSO National Laboratories in Singapore for funding and A.B. thanks the Royal Society and the GCHQ for support through the James Ellis University Research Fellowship. This work was supported in part by the UK Engineering and Physical Sciences Research Council through its interdisciplinary Research Collaboration on Quantum Information Processing, and by the European Union Networks QGATES and SCALA.

\end{document}